\documentclass[aps,prl,reprint,twocolumn,superscriptaddress,floatfix,nofootinbib,longbibliography]{revtex4-2}
\usepackage{graphicx,amsmath,amsfonts,amssymb,amsthm,xr}
\usepackage{epsfig,amsmath,amssymb,color,dsfont,upgreek,physics}
\usepackage{mathrsfs}
\usepackage{mathtools}
\usepackage{bbold}
\usepackage{float}
\usepackage[caption=false]{subfig}

\usepackage[bookmarks=true,colorlinks,linkcolor=OrangeRed,urlcolor=NavyBlue,citecolor=RoyalBlue]{hyperref}
\usepackage[dvipsnames]{xcolor}
\usepackage{orcidlink}

\usepackage{graphicx}
\usepackage{dcolumn}
\usepackage{bm}
\usepackage{color}
\newcommand{\Zt}{\mathbb{Z}_2}
\newcommand{\rr}{\mathbf{r}}

\begin{document}

\title{Role of Plaquette Term in Genuine $2+1$D String Dynamics on Quantum Simulators}

\author{Yizhuo Tian}
\thanks{These authors contributed equally to this work.}
\affiliation{Department of Physics and Arnold Sommerfeld Center for Theoretical Physics (ASC), Ludwig Maximilian University of Munich, 80333 Munich, Germany}

\author{N.~S.~Srivatsa}
\thanks{These authors contributed equally to this work.}
\affiliation{Max Planck Institute of Quantum Optics, 85748 Garching, Germany}
\affiliation{Munich Center for Quantum Science and Technology (MCQST), 80799 Munich, Germany}

\author{Kaidi Xu${}^{\orcidlink{0000-0003-2184-0829}}$}
\affiliation{Max Planck Institute of Quantum Optics, 85748 Garching, Germany}
\affiliation{Munich Center for Quantum Science and Technology (MCQST), 80799 Munich, Germany}

\author{Jesse J.~Osborne${}^{\orcidlink{0000-0003-0415-0690}}$}
\affiliation{Max Planck Institute of Quantum Optics, 85748 Garching, Germany}
\affiliation{Munich Center for Quantum Science and Technology (MCQST), 80799 Munich, Germany}

\author{Umberto Borla${}^{\orcidlink{0000-0002-4224-5335}}$}
\affiliation{Max Planck Institute of Quantum Optics, 85748 Garching, Germany}
\affiliation{Munich Center for Quantum Science and Technology (MCQST), 80799 Munich, Germany}

\author{Jad C.~Halimeh${}^{\orcidlink{0000-0002-0659-7990}}$}
\email{jad.halimeh@lmu.de}
\affiliation{Max Planck Institute of Quantum Optics, 85748 Garching, Germany}
\affiliation{Department of Physics and Arnold Sommerfeld Center for Theoretical Physics (ASC), Ludwig Maximilian University of Munich, 80333 Munich, Germany}
\affiliation{Munich Center for Quantum Science and Technology (MCQST), 80799 Munich, Germany}

\date{\today}

\begin{abstract}
With the advent of quantum simulators of $2+1$D lattice gauge theories (LGTs), a fundamental open question is under what circumstances the observed physics is genuinely $2+1$D rather than effectively $1+1$D. Here, we address this question in the ongoing strong effort to quantum-simulate string dynamics in $2+1$D LGTs on state-of-the-art quantum hardware. Through tensor network simulations and analytic derivations, we show that the plaquette term, which represents a magnetic field and only emerges in $d>1$ spatial dimensions, plays a crucial role in \textit{genuine} $2+1$D string dynamics deep in the confined regime. In its absence and for minimal-length (Manhattan-distance) strings, we demonstrate how string breaking, although on a lattice in $d=2$ spatial dimensions, can be effectively mapped to a $1+1$D dynamical process independently of lattice geometry. Our findings not only answer the question of what qualifies as genuine $2+1$D string dynamics, but also serve as a clear guide for future quantum simulation experiments of $2+1$D LGTs.
\end{abstract}

\maketitle

\textbf{\emph{Introduction.---}}String breaking is a quintessential phenomenon in high-energy physics (HEP) with direct connections to quantum chromodynamics (QCD). Pulling apart a quark-antiquark pair beyond a certain distance leads to the flux string between them becoming so energetically expensive that new quark-antiquark pairs are created to break it \cite{Weinberg_book,Gattringer_book,Zee_book}. Studying this and other phenomena in $3+1$D QCD is an outstanding challenge and at the heart of ongoing efforts at dedicated particle colliders such as the LHC and RHIC \cite{Ellis_book}. These experiments are complemented by Monte Carlo (MC) techniques applied to lattice QCD, a well-defined UV-regulated version of the theory on a lattice. These techniques, particularly in the Euclidean path integral formulation, have allowed accurate estimations of, e.g., decay constants, hadron masses, and thermodynamic properties of QCD \cite{Creutz1980MC,Montvay_book}. However, MC methods are encumbered by the infamous sign problem \cite{Troyer2005computational} when it comes to studying out-of-equilibrium processes such as dynamical string breaking.

Lattice gauge theories (LGTs) have emerged as a powerful tool to study the real-time dynamics of HEP phenomena such as string breaking \cite{Kogut1975,Rothe_book}, in addition to their pivotal role in investigating the equilibrium physics of HEP models through MC techniques \cite{Creutz1980MC}, in offering descriptions of exotic phases in condensed matter systems \cite{Wegner1971,Kogut_review,wen2004quantum,Savary2016,Calzetta_book}, as well as in hosting intriguing nonergodic quantum many-body phases \cite{Smith2017,Brenes2018,smith2017absence,karpov2021disorder,Sous2021,Chakraborty2022,Halimeh2021enhancing,Surace2020,Lang2022SGP,Desaules2022weak,Desaules2022prominent,aramthottil2022scar,Tarabunga2023many,Hartse:2024qrv,Desaules2024ergodicitybreaking,desaules2024massassistedlocaldeconfinementconfined,Jeyaretnam2025Hilbert,Smith2025nonstabilizerness,Falcao2025Nonstabilizerness,Esposito2025magic,ciavarella2025generichilbertspacefragmentation,ciavarella2025truncationuncertaintiesaccuratequantum,steinegger2025geometricfragmentationanomalousthermalization}. The real-time dynamics of LGTs has been extensively studied using tensor networks \cite{Montangero_book,Uli_review,Paeckel_review,Orus2019,Banuls_review,Banuls2018} and, more recently, quantum simulators \cite{Georgescu_review,Bloch2008,Gross2017,Alexeev_review}. In the context of LGTs, the latter have emerged as a complementary venue that can provide a first-principles study of the dynamical processes involved in HEP phenomena, while also affording a quantum advantage that can go beyond what is accessible through numerical methods such as tensor networks. The quantum simulation of LGTs is currently an extremely active field \cite{Dalmonte_review, Zohar_review, aidelsburger2021cold, Zohar_NewReview, klco2021standard,Bauer_ShortReview, Bauer_review, dimeglio2023quantum, Cheng_review, Halimeh_review, Cohen:2021imf, Lee:2024jnt, Turro:2024pxu,bauer2025efficientusequantumcomputers}, and the last decade has seen an impressive suite of experiments observing various phenomena relevant to both HEP and quantum many-body physics \cite{Martinez2016,Klco2018,Goerg2019,Schweizer2019,Mil2020,Yang2020,Wang2021,Su2022,Zhou2022,Wang2023,Zhang2023,Ciavarella2024quantum,Ciavarella:2024lsp,Farrell:2023fgd,Farrell:2024fit,zhu2024probingfalsevacuumdecay,Ciavarella:2021nmj,Ciavarella:2023mfc,Ciavarella:2021lel,Gustafson:2023kvd,Gustafson:2024kym,Lamm:2024jnl,Farrell:2022wyt,Farrell:2022vyh,Li:2024lrl,Zemlevskiy:2024vxt,Lewis:2019wfx,Atas:2021ext,ARahman:2022tkr,Atas:2022dqm,Mendicelli:2022ntz,Kavaki:2024ijd,Than:2024zaj,Angelides2025first,cochran2024visualizingdynamicschargesstrings,gyawali2024observationdisorderfreelocalizationefficient,gonzalezcuadra2024observationstringbreaking2,crippa2024analysisconfinementstring2,schuhmacher2025observationhadronscatteringlattice,davoudi2025quantumcomputationhadronscattering,cobos2025realtimedynamics21dgauge,saner2025realtimeobservationaharonovbohminterference}.

\begin{figure}[t!]
\includegraphics[width=\linewidth]{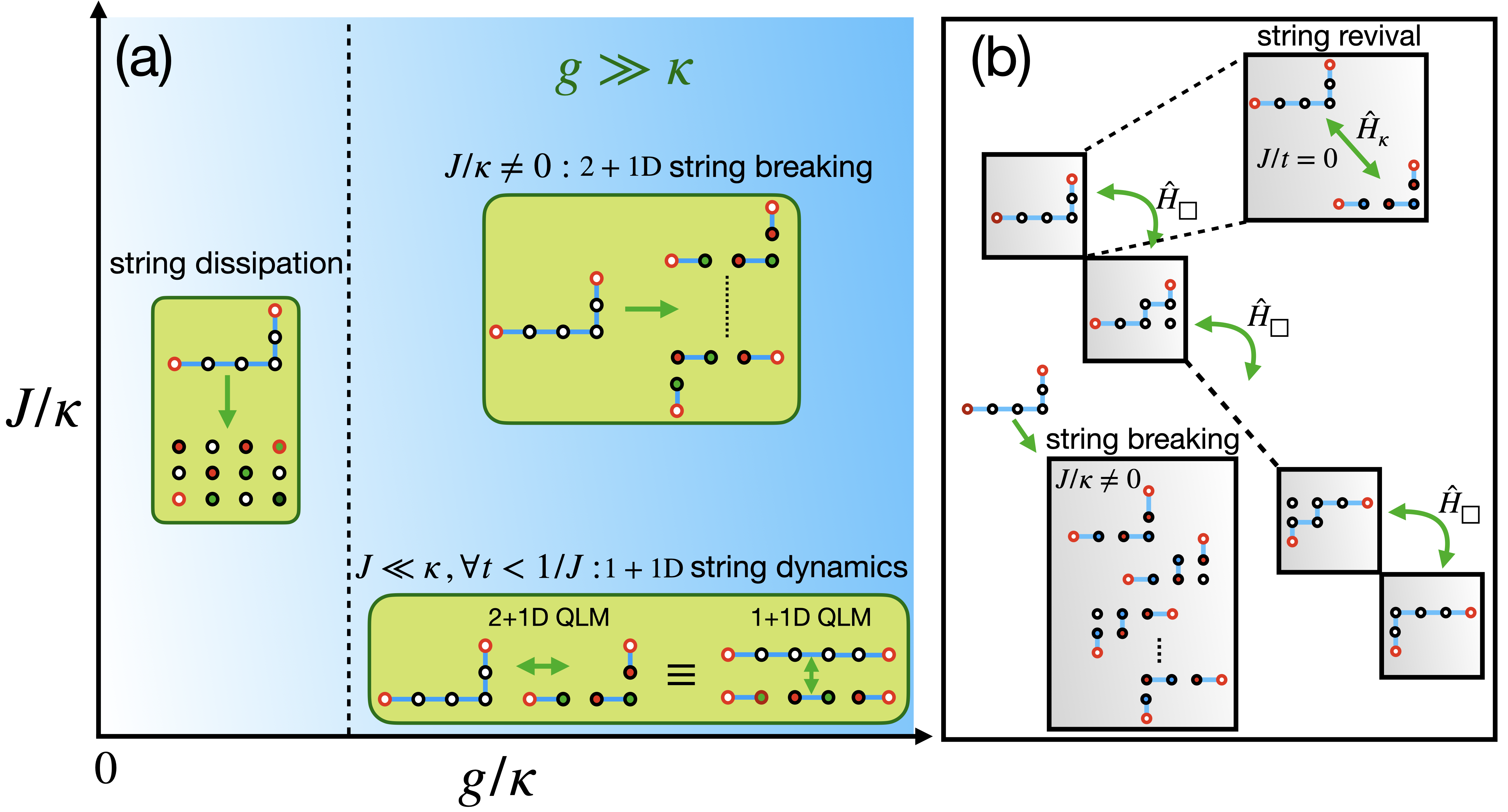}
\caption{(a) Schematic describing the string dynamics at resonance ($2m=g$) in different regimes of the $2+1$D U$(1)$ quantum link model (QLM). In the confined regime ($g\gg \kappa$) and for $J\ll \kappa$, we expect string dynamics to be $1+1$D for all times $t<1/J$ and map to the $1+1$D QLM, while for $J/\kappa\neq0$, we expect string breaking. In the regime where $g\ll \kappa$, we expect string dissipation through a trivial thermalization process leading to matter creation everywhere in the lattice. (b) Schematic for the minimal model, which we define in the main text. All unbroken string configurations are connected through the plaquette term $\hat{H}_{\Box}$, while broken string configurations are accessed via the hopping term $\hat{H}_\kappa$. At resonance, the effective string dynamics is restricted to the manifold spanned by these states.}
\label{fig:phase}
\end{figure}

In particular, several experiments have recently appeared that probe string dynamics on quantum simulators of $2+1$D LGTs \cite{cochran2024visualizingdynamicschargesstrings,gonzalezcuadra2024observationstringbreaking2,cobos2025realtimedynamics21dgauge}, complemented by tensor network studies \cite{borla2025stringbreaking21dmathbbz2,xu2025tensornetworkstudyrougheningtransition,dimarcantonio2025rougheningdynamicselectricflux,xu2025stringbreakingdynamicsglueball}. In terms of accessible physics, LGTs in $d=2$ spatial dimensions go fundamentally beyond their $d=1$ counterparts in large part because of the presence of a magnetic field in the form of a plaquette term, which is absent in one spatial dimension. Although the \texttt{Google Quantum AI} experiment \cite{cochran2024visualizingdynamicschargesstrings} includes a plaquette term in a $\mathbb{Z}_2$ LGT on a square lattice, the \texttt{QuEra} experiment of a U$(1)$ LGT \cite{gonzalezcuadra2024observationstringbreaking2} and the \texttt{IBM} experiment of a $\mathbb{Z}_2$ LGT \cite{cobos2025realtimedynamics21dgauge}, both on a hexagonal lattice, do not. A common claim, repeated in Ref.~\cite{cobos2025realtimedynamics21dgauge}, is that a perturbative plaquette term emerges at $n^\text{th}$ order in the minimal coupling (with $n$ the number of links in the plaquette) and compensates for the absence of an explicit plaquette term. However, this has hitherto not been thoroughly investigated despite the important possible implications.

In this Letter, we ask: Under what circumstances is string breaking in $2+1$D LGTs genuinely a $2+1$D process? We answer this question through extensive tensor network simulations---supported by analytic derivations---of string dynamics and breaking in two paradigmatic U$(1)$ and $\mathbb{Z}_2$ LGTs in on square and hexagonal lattices. We find that an explicit plaquette term is necessary for genuine $2+1$D string dynamics. Indeed, we demonstrate that string breaking in a $2+1$D LGT maps to a $1+1$D process in the absence of a plaquette term. We also show that the perturbative plaquette term arising from higher orders of the minimal coupling is insufficient to induce genuine $2+1$D behavior, as it is strongly suppressed in the confined regime. In contrast, we show that in the presence of the plaquette term a rich genuine $2+1$D string dynamics occurs where the wave function explores a subspace of (un)broken string configurations; see Fig.~\ref{fig:phase}.

\textbf{\emph{Model.---}}Even though our results are general, we first focus on a $2+1$D U$(1)$ quantum link model (QLM) \cite{Chandrasekharan1997,Wiese_review,Kasper2017} defined on a square lattice given by the Hamiltonian
\begin{align}\nonumber
\hat{H}=&\underbrace{-\kappa\sum_{\mathbf{j},\mu}\Big(s_{\mathbf{j},\mathbf{e}_{\mu}}\hat{\phi}^{\dagger}_{\mathbf{j}}\hat{U}_{\mathbf{j},\mathbf{e}_{\mu}}\hat{\phi}_{\mathbf{j}+\mathbf{e}_{\mu}}+\textrm{h.c.}}_{\hat{H}_\kappa}\Big)+m\sum_{\mathbf{j}}s_{\mathbf{j}}\hat{\phi}^{\dagger}_{\mathbf{j}}\hat{\phi}_{\mathbf{j}}\\
&+g\sum_{\mathbf{j},\mu}\hat{S}_{\mathbf{j},\mathbf{e}_{\mu}}^z-\underbrace{J\sum_{\Box}\Big(\hat{U}_{\Box}+\hat{U}^{\dagger}_{\Box}\Big)}_{\hat{H}_{\Box}}.
\label{eq:hamiltonian}
\end{align} 
The first term describes the minimal coupling between the matter fields $\hat{\phi}_\mathbf{j}$ representing hardcore bosons~\cite{hcb} residing on the lattice sites $\mathbf{j} = (j_x, j_y)^\intercal$ and the gauge fields $\hat{U}_{\mathbf{j},\mathbf{e}_{\mu}}$, which are defined on the links connecting sites $\mathbf{j}$ and $\mathbf{j} + \mathbf{e}_\mu$. To restrict the Hilbert space of the gauge fields, we adopt the spin-$\tfrac{1}{2}$ representation, which provides the simplest nontrivial realization compatible with current quantum simulation platforms \cite{gonzalezcuadra2024observationstringbreaking2}. In this mapping, the gauge field operator is given by $\hat{U}_{\mathbf{j},\mathbf{e}_\mu} = \hat{S}^+_{\mathbf{j},\mathbf{e}_\mu}$, and the corresponding local electric field is represented by $\hat{S}^z_{\mathbf{j},\mathbf{e}_\mu}$. The second term sets the staggered mass of the matter fields. This formulation adopts the Kogut--Susskind prescription for staggered fermions, resulting in both the hopping and mass terms being staggered \cite{Kogut_review}. The staggering in the hopping term is direction-dependent with  $s_{\mathbf{j},\mathbf{e}_x} = +1$ and $s_{\mathbf{j},\mathbf{e}_y} = (-1)^{j_x}$. For the mass term, the staggering is given by $s_{\mathbf{j}} = (-1)^{j_x + j_y}$, such that a particle located on an even site ($s_{\mathbf{j}} = +1$) corresponds to a positive charge, while its absence on an odd site ($ s_{\mathbf{j}} = -1$) corresponds to a negative charge. The third term is a linear electric field with strength $g$ which we use to introduce string tension in the model since the natural electric field energy is trivial in our model once we restrict to spin-$1/2$, as then $\hat{E}_{\mathbf{j},\mathbf{e}_{\mu}}^2=(\hat{S}^z_{\mathbf{j},\mathbf{e}_\mu})^2=\hat{\mathds{1}}/4$. The final term represents the magnetic energy via the plaquette interaction $\hat{U}_\Box = \hat{U}_{\mathbf{j},\mathbf{e}_x} \hat{U}_{\mathbf{j}+\mathbf{e}_x,\mathbf{e}_y} \hat{U}^\dagger_{\mathbf{j}+\mathbf{e}_y,\mathbf{e}_x} \hat{U}^\dagger_{\mathbf{j},\mathbf{e}_y}$, which contributes only for oriented plaquettes that can be flipped~\cite{Hashizume2022}. The Hamiltonian is invariant under local gauge transformations generated by $\hat{G}_{\mathbf{j}} = \hat{\phi}^{\dagger}_{\mathbf{j}} \hat{\phi}_{\mathbf{j}} - \sum_{\mu}\big( \hat{E}_{\mathbf{j},\mathbf{e}_{\mu}} - \hat{E}_{\mathbf{j}-\mathbf{e}_{\mu},\mathbf{e}_{\mu}}\big)-\big[1 - (-1)^{\mathbf{j}}\big]/2$.

\begin{figure}

  \includegraphics[width=\linewidth]{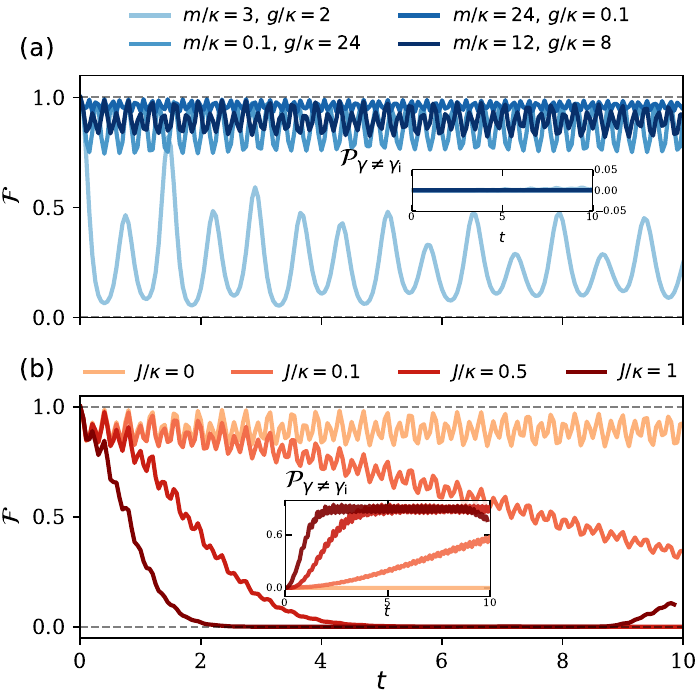}
  \caption{Dynamics of the initial-state fidelity following quantum quenches of an L-shaped string for different off-resonant parameters in the $2+1$D QLM, calculated using MPS numerics. The insets show the overlap  $\mathcal{P}_{\gamma\neq\gamma_{\text{i}}}$ between the time-evolved string and all other string configurations that have the same Manhattan distance as the L-shaped string. (a) We fix $J/\kappa=0$ and choose different values for $m/\kappa$ and $g/\kappa$. (b) We fix $m/\kappa=12,\,g/\kappa=8$ and choose different values of $J/\kappa$.}
  \label{fourth_order}
\end{figure}

\textbf{\emph{String dynamics.---}}We investigate the dynamics of minimal-length strings connecting two static charges. This restriction to minimal-length strings (in Manhattan distance) naturally arises in the QLM with gauge fields represented by spin-$1/2$ operators, since non-minimal strings must bend around corners, which leads to a violation of Gauss's law---more details are provided in the Supplemental Material (SM) \cite{SM}. 
The initial configuration is prepared by placing negative and positive static charges on an even (\(\mathbf{j}_\text{e}\)) and an odd (\(\mathbf{j}_\text{o}\)) site, respectively, and connecting them by a string of oriented electric field lines. The sites containing the charges satisfy the local constraints $\hat{G}_{\mathbf{j}_{\textrm{e/o}}}|\Psi\rangle = \pm 1$, while at all other sites $\hat{G}_{\mathbf{j}}|\Psi\rangle = 0$. We consider a cylinder with dimensions \( L_x = 7 \) and \( L_y = 6 \), and place two charges at the opposite corners of a patch of size \( l_1 \times l_2 \), with \( l_1 = 5 \) and \( l_2 = 4\) and connect them with a string of minimal length featuring an L-shape; see Fig.~\ref{fig:phase} and the SM \cite{SM} for specifics. We numerically study the dynamics of the string using the time-dependent variational principle (TDVP) algorithm~\cite{Haegeman2011,Haegeman2013,Haegeman2016} implemented in the \texttt{Matrix Product Toolkit} \cite{mptoolkit}. To ensure convergence, we use matrix product state (MPS) bond dimension $\chi=256$ and TDVP step size $\delta t=0.01$ for all our computations \cite{SM}. 

To see clear string dynamics, one must be in the confined phase where spurious matter creation outside the string configuration is suppressed. This is true when $2m+g\gg \kappa$, where $2m+g$ is the energy needed to create a particle-antiparticle pair from the vacuum. String breaking takes place when the energy stored in the string matches the energy required to create a dynamical particle-antiparticle pair. In the QLM with $S=1/2$, the energy stored in a string segment consisting of $n$ links is $ng/2$, where $g/2$ is the energy contribution per link. For an $n$\textsuperscript{th}-order process, breaking the string at that segment involves flipping $n$ links and creating a particle-antiparticle pair at the ends of this segment. The energy of the broken segment is $2m - ng/2$. Setting this equal to the energy stored in the unbroken segment yields the resonance condition:
$2m = ng$. It is important to note that, within the staggered fermion framework, the number of links $n$ in a breakable string segment is always an odd integer. 

To probe string dynamics and its breaking, we examine the fidelity to the initial string state $\mathcal{F}(t)=\lvert\braket{\psi_0}{\psi(t)}\rvert^2$, the overlap $\mathcal{P}_{\gamma\neq\gamma_{\text{i}}}=\sum_{\gamma\neq\gamma_{\text{i}}} \lvert\braket{\psi_\gamma}{\psi(t)}\rvert^2$ with the set of all minimal string configurations $\gamma$ excluding the initial string $\gamma_{\text{i}}$, and the total matter occupation within the patch, \(\langle \hat{n}\rangle = \sum_{\mathbf{j} \in \mathrm{sites}} \langle \hat{n}_{\mathbf{j}} \rangle \), with the staggered fermion number operator defined as $\hat{n}_{\mathbf{j}} = (-1)^\mathbf{j} \big\{\hat{\phi}^{\dagger}_{\mathbf{j}} \hat{\phi}_{\mathbf{j}} - \big[1 - (-1)^\mathbf{j}\big]/2 \big\}$.   

\begin{figure}
\includegraphics[width=\linewidth]{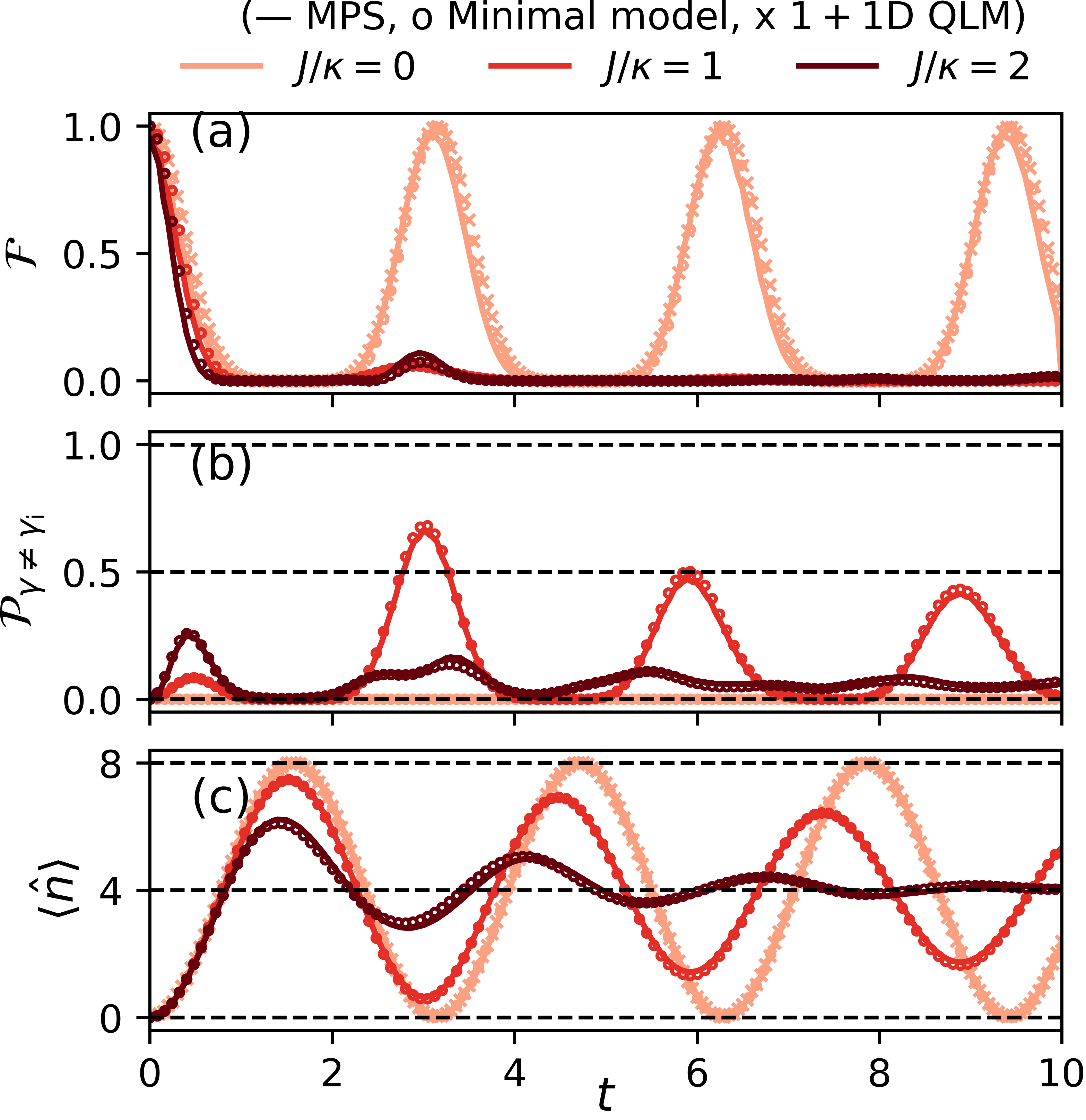}
\caption{Resonant string dynamics at $m/\kappa=12$, $g/\kappa=24$ and different values of $J/\kappa$ following the evolution of an L-shaped string in the $2+1$D QLM computed using MPS-based simulations, and compared with the minimal model~$\hat{H}_\text{min}$.  (a) The fidelity $\mathcal{F}$ with the initial string state. (b) The total overlap $\mathcal{P}_{\gamma\neq\gamma_{\text{i}}}$ with all minimal strings excluding the initial string. (c) The total matter occupation $\langle\hat{n}\rangle$ computed within the minimal patch containing the two static charges.  For $J/\kappa=0$, the $1+1$D QLM data is also shown for total matter occupation and the fidelity.} 
\label{reson}
\end{figure}

The primary question that we are concerned with is whether the string dynamics remains genuinely $2+1$D in the absence of explicit plaquette terms. Although an effective plaquette interaction can arise in perturbation theory in fourth order of the minimal coupling,  this is expected to be strongly suppressed in the confined regime, and its effects are expected to manifest only at very long timescales \( t \sim g^3 / \kappa^4 \). We numerically investigate this by turning off the plaquette term, \( J/\kappa = 0 \), and choose parameters away from resonance to prevent undesired string breaking. We then determine whether the string evolves into a configuration different from the one it started with. We show in Fig.~\ref{fourth_order}(a) that the hopping process alone is \emph{incapable} of inducing string dynamics beyond the initial string configuration in the confined phase, which is evidenced by the nonzero fidelity $\mathcal{F}$ and vanishing $\mathcal{P}_{\gamma\neq\gamma_{\text{i}}}$ (shown in the inset). However, with nonzero plaquette terms ($J/\kappa\neq 0$), we show in Fig.~\ref{fourth_order}(b) that the dynamics is now $2+1$D, as we can observe nontrivial string dynamics, as evidenced by the decaying fidelity and nonzero $\mathcal{P}_{\gamma\neq\gamma_{\text{i}}}$ (inset).

At the first-order resonance (\( 2m = g \)), and in the absence of the magnetic term (\( J/\kappa = 0 \)), the dynamics remains restricted along the initial string as long as one is deep in the confined phase. As shown in Fig.~\ref{reson}, we observe near-perfect revivals in the fidelity $\mathcal{F}$ and the total matter occupation $\langle \hat{n}\rangle$ within the minimal patch. We also see that $\mathcal{P}_{\gamma\neq\gamma_{\text{i}}}$ is zero, confirming that the dynamics is effectively $1+1$D. Indeed, this behavior is nearly identical to the dynamics of a string in the $1+1$D QLM \cite{qlm1d}, which we demonstrate by comparing the $2+1$D dynamics with a simulation of the $1+1$D QLM in Fig.~\ref{reson}. However, once the magnetic term is turned on (\( J/\kappa \neq 0 \)), the revivals become increasingly suppressed, eventually resulting in permanent string breaking in the \( t \to \infty \) limit. In this regime, the system explores various broken string configurations within the same spatial region. This behavior is reflected in the decay of fidelities and suppression of revivals in \( \mathcal{P}_{\gamma\neq\gamma_{\text{i}}} \), accompanied by the saturation of the matter occupation to a nonzero value, which is consistent with the production of additional charge pairs enabling \textit{genuine} \( 2+1 \)D string breaking. An important observation is that increasing \(J/\kappa\) leads to faster dynamics and enhances dissipation within the string subspace. This facilitates the early exploration of a broader range of minimal string configurations by the wave function, as evidenced by the prominent peak in \( \mathcal{P}_{\gamma\neq\gamma_{\text{i}}} \) at early time for \(J/\kappa=2\). The string breaks at later times, and consequently  \( \mathcal{P}_{\gamma\neq\gamma_{\text{i}}} \) does not return to high values, in contrast to the case with  \(J/\kappa=1\). We illustrate the resonant string dynamics for the aforementioned regimes schematically in Fig.~\ref{fig:phase}(a). We also provide videos for both off- and on-resonant string dynamics with(out) the plaquette term \cite{videos}. 

\textbf{\emph{Minimal model.---}}To capture the resonant string breaking dynamics deep in the confined regime, we construct a minimal Hamiltonian $\hat{H}_{\textrm{min}} = \hat{P} \hat{H} \hat{P}$ where \(\hat{P}\) is the projector onto a fixed-energy manifold comprising both unbroken and broken string states within the region connecting the static charges. All unbroken string configurations can be generated by repeated applications of the plaquette term \(\hat{H}_{\Box}\) on a given initial string state. Similarly, starting from a minimal string configuration, all broken string states can be accessed via the hopping term \(\hat{H}_\kappa\). More details of the construction can be found in the SM \cite{SM}. In the absence of the plaquette term (\(J/\kappa = 0\)), the evolution is entirely confined to the subspace spanned by broken and unbroken string states derived from the initial string. Unremarkably, this restricted dynamics leads to perfect string revivals. When the plaquette term is activated (\(J/\kappa \neq 0\)), additional broken string configurations become accessible beyond the subspace of the initial string. This can be understood as an effective dissipation process, wherein the initial string dephases into the rest of the string-configuration subspace within the patch, resulting in genuine $2+1$D string breaking. This is depicted in Fig.~\ref{fig:phase}(b). As shown in Fig.~\ref{reson}, we observe excellent agreement between this effective model and full MPS simulations for small values of \(J/\kappa\), validating the description of string dynamics at resonance within the projected subspace.

\begin{figure}
\includegraphics[width=\linewidth]{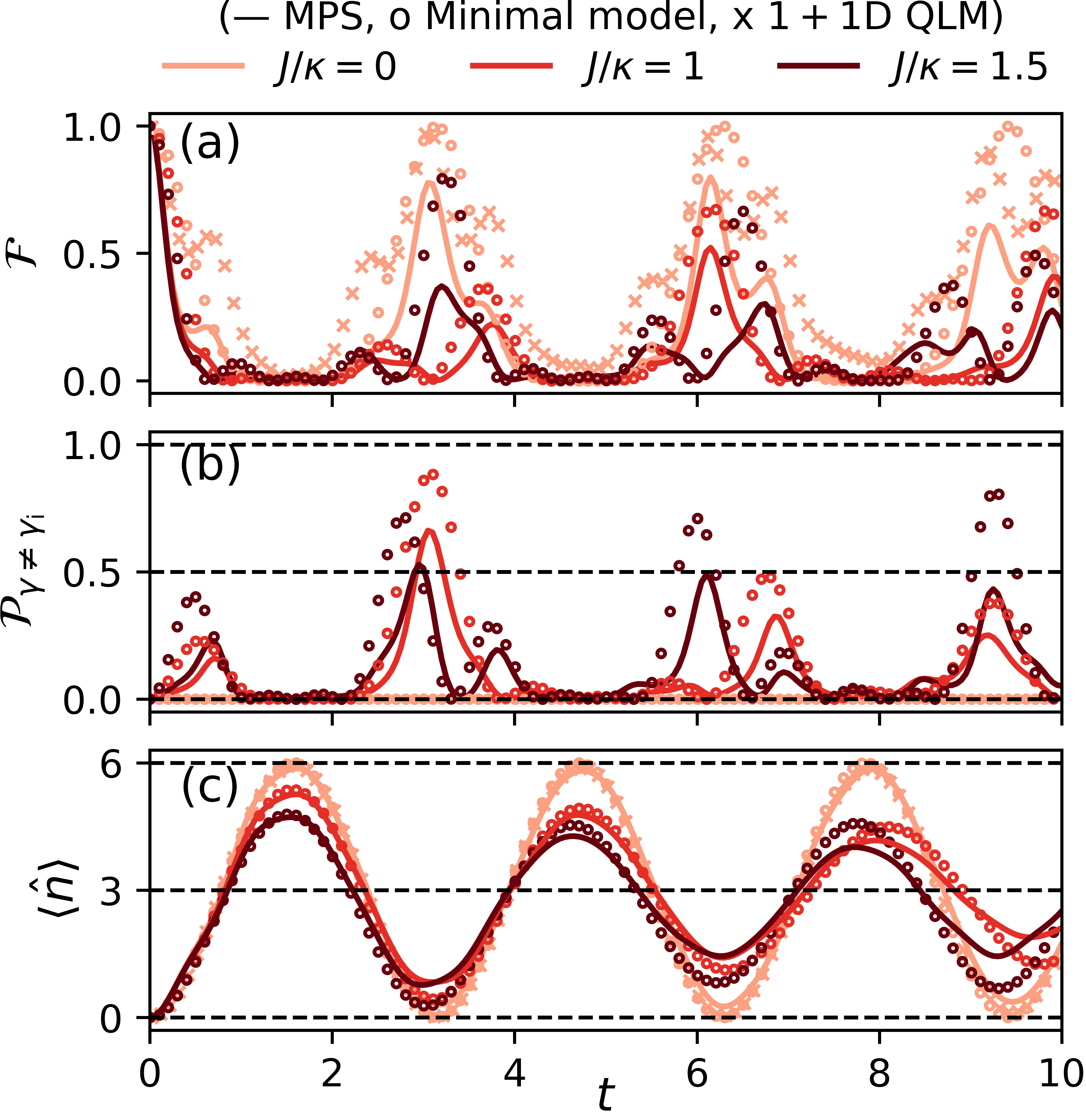}
\caption{Resonant string dynamics at $m/\kappa=2$, $g/\kappa=4$ and different values of $J/\kappa$ computed using MPS-based simulations and compared with the minimal model $\hat{H}_{\mathrm{min}}$ for the hexagonal QLM. (a) The fidelity $\mathcal{F}$ with the initial string state. (b) The total overlap $\mathcal{P}_{\gamma \neq \gamma_{\text{i}}}$ with all minimal strings excluding the initial string. (c) The total matter occupation $\langle\hat{n}\rangle$ computed within the minimal patch containing the two static charges. For $J/\kappa=0$, the $1+1$D QLM data is also shown for total matter occupation and the fidelity.} 
\label{reson_hex}
\end{figure}

\textbf{\emph{Effects of lattice geometry.---}}It is natural to ask whether the underlying lattice geometry influences the dimensionality of string dynamics. To investigate this, we consider a QLM defined on an experimentally relevant hexagonal lattice as in the \texttt{QuEra} experiment \cite{gonzalezcuadra2024observationstringbreaking2}, and study string dynamics at resonance both with and without the plaquette terms. Details about the Hamiltonian and the initial string state that we consider for this geometry can be found in the SM \cite{SM}. Owing to the hexagonal geometry, the effective plaquette term generated via virtual hopping is even more suppressed in the confined phase, as it arises only at sixth order of perturbation theory. Here we repeat our analysis of string dynamics, following the same approach used for the square lattice. We investigate the resonance condition by choosing parameters similar to those used in the \texttt{QuEra} experiment \cite{gonzalezcuadra2024observationstringbreaking2}. As shown in Fig.~\ref{reson_hex}, for the of case $J/\kappa=0$, we observe revivals in the matter occupation within the minimal patch, $\langle\hat{n}\rangle$, agreeing well with $1+1$D QLM and the minimal model analysis, similar to the case of the square lattice. Likewise, the probability to visit other minimal string configurations, $P_{\gamma\neq\gamma_{\text{i}}}$, remains vanishing, indicating $1+1$D string dynamics. For $J/\kappa\neq 0$, we see a suppression of revivals in the matter occupation and a corresponding drop in fidelity, along with nonzero $P_{\gamma\neq\gamma_{\text{i}}}$, indicative of genuine $2+1$D string breaking. Interestingly, the suppression of revivals in the matter occupation is less pronounced than in the case of the square lattice. The difference arises from the hexagonal geometry, which supports fewer minimal string configurations (that are accessible via applications of the plaquette term) for a given Manhattan distance compared to the square lattice. This also explains why the revivals in the hexagonal geometry are robust even for the modest choice of $m/\kappa,\;g/\kappa$ we have made. It is important to note that the computed fidilities between the MPS numerics and the $1+1$D QLM show limited agreement. The discrepancy arises because, in the parameter regimes considered, the matter and gauage fluctuations outside the initial string behave differently due to difference in dimensionality. Consequently, we also do not expect a strong agreement with the minimal model analysis in the parameter regime considered.  Details on off-resonant dynamics can be found in the SM \cite{SM}.

\textbf{\textit{Summary and outlook.---}}In this work, we have focused on string dynamics in the deeply confined phase of $2+1$D LGTs. In this phase, strings are well-defined and resonance conditions facilitate their breaking. We have investigated the effect of the plaquette term on the $2+1$D nature of string dynamics. Through tensor network simulations and analytic arguments, we have shown that a plaquette term is necessary for genuine $2+1$D behavior. In its absence, string breaking is effectively a $1+1$D dynamical process. For our main findings, we have focused on a $2+1$D U$(1)$ QLM both on a square and a hexagonal lattice, highlighting the independence of our conclusions on the lattice geometry, and also bringing our results to direct relevance to recent experimental work \cite{gonzalezcuadra2024observationstringbreaking2}.

Our conclusions are general and apply to other gauge groups such as $\mathbb{Z}_2$ and hold for different initial states, as we detail in the SM \cite{SM}. Our findings set a benchmark for what constitutes genuine $2+1$D string dynamics and breaking, showing how the physics of $2+1$D LGTs can still be effectively $1+1$D in the absence of a magnetic field, and provide a guide for future quantum simulation experiments of LGTs in higher spatial dimensions. A direct followup of our work would be to consider other phenomena in $2+1$D LGTs and analyze the effect of the (absence of the) plaquette term on them when it comes to genuine $2+1$D dynamics. These can include HEP phenomena such as scattering \cite{su2024particlecollider,schuhmacher2025observationhadronscatteringlattice,davoudi2025quantumcomputationhadronscattering}, but also condensed matter dynamics such as disorder-free localization \cite{gyawali2024observationdisorderfreelocalizationefficient}.

\bigskip

\footnotesize
\begin{acknowledgments}
    The authors acknowledge funding by the Max Planck Society, the Deutsche Forschungsgemeinschaft (DFG, German Research Foundation) under Germany’s Excellence Strategy – EXC-2111 – 390814868, and the European Research Council (ERC) under the European Union’s Horizon Europe research and innovation program (Grant Agreement No.~101165667)—ERC Starting Grant QuSiGauge. Views and opinions expressed are however those of the author(s) only and do not necessarily reflect those of the European Union or the European Research Council Executive Agency. Neither the European Union nor the granting authority can be held responsible for them. This work is part of the Quantum Computing for High-Energy Physics (QC4HEP) working group.
\end{acknowledgments}
\normalsize

\bibliography{biblio}

\begin{thebibliography}{120}
\bibitem [{qlm()}]{qlm1d}%
  \BibitemOpen
  \href@noop {} {}\bibinfo {note} {For the $1+1$D QLM, the Hamiltonian takes
  the same form as Eq.~1 in the main text, but without the plaquette term
  (i.e., $J = 0$). To enable a meaningful comparison with string dynamics in
  the $2+1$D QLM, we construct a string of comparable length by flipping links
  in 1-D vacuum configuration.}\BibitemShut {Stop}%
\end{thebibliography}

\pagebreak
\clearpage
\widetext
\begin{center}
\textbf{\large Supplemental Material for `Role of Plaquette Term in Genuine $2+1$D String Dynamics on Quantum Simulators'}
\end{center}
\makeatletter
\renewcommand{\c@secnumdepth}{0}
\makeatother
\setcounter{equation}{0}
\setcounter{figure}{0}
\setcounter{table}{0}
\setcounter{page}{1}
\makeatletter
\renewcommand{\theequation}{S\arabic{equation}}
\renewcommand{\thefigure}{S\arabic{figure}}
\renewcommand{\thesection}{S\arabic{section}}

\section{$\mathbb{Z}_2$ lattice gauge theory}
We address here whether the results obtained in the main text apply to the related $\mathbb{Z}_2$ lattice gauge theory with Ising matter. For completeness, we consider both square and hexagonal lattices, as they are relevant to different experimental setups. The Hamiltonian of this model is

\begin{align}
\hat{H}=&-m \sum_\rr \hat A_{\rr} -J \sum_{\rr^*} \hat B_{\rr^*} -\kappa \sum_{\rr,\eta} \hat\sigma^z_{\rr,\eta} -g \sum_{\rr,\eta} \hat\sigma^x_{\rr,\eta},
\label{eq:2DFS}
\end{align}
where, compared to the standard literature, the names of the couplings have been adapted to stress the analogy with Eq. \eqref{eq:hamiltonian}. Here $\hat{A}_{\rr}$ is the product of $\hat\sigma^x$ on the links connecting at the vertex $\rr$, while $\hat{B}_{\rr}$ is the plaquette operator, i.e., the product of $\sigma^z$ around a minimal closed loop. Note how these two definitions do not depend on the specific lattice, and can be adapted to both the square and hexagonal lattices that we address here. 

\subsection{Hexagonal lattice}
\begin{figure*}[ht]
 \centering
 \includegraphics[width=\textwidth]{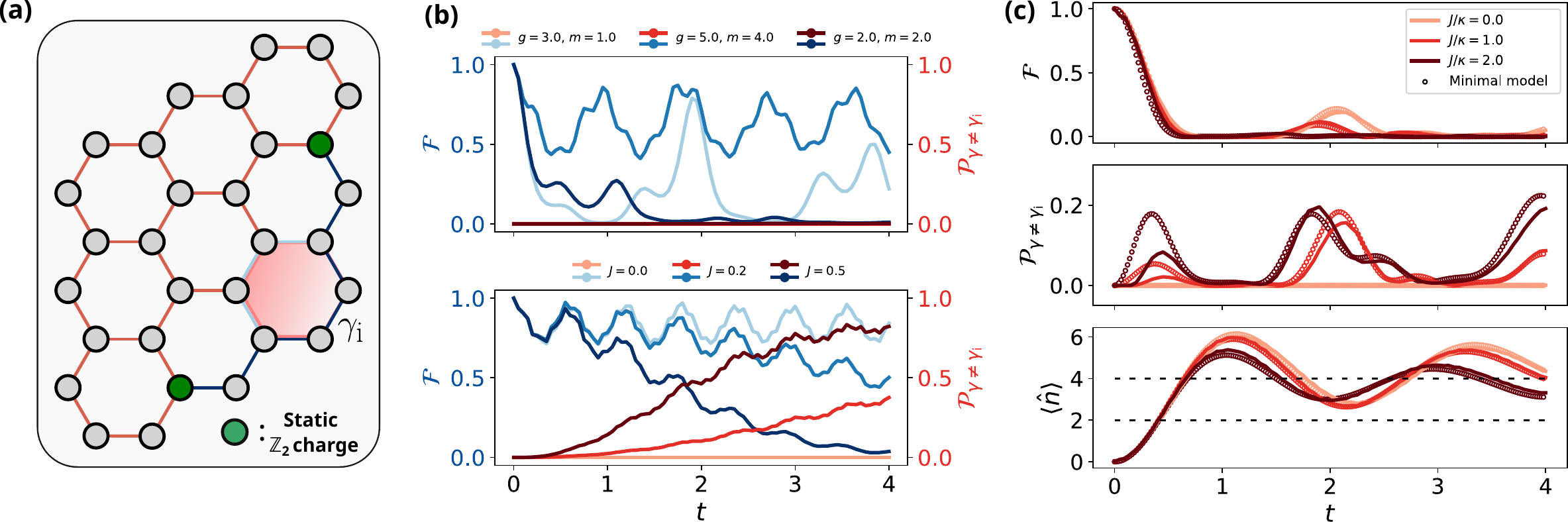}
 \caption{TDVP results for the time evolution of the string depicted in (a) under quenches at values of $g/m$ away (b) and in correspondence (c) of the $1^{\rm st}$ order resonance $g=2m$. In each plot, we test whether $2+1$D string dynamics occurs by measuring the transition rate to the other minimal string configurations $\mathcal{P_{\gamma \neq \gamma_{\rm i}}}$ together with the fidelity $\mathcal{F}$ with the initial configuration. In both cases, in the absence of the plaquette term $J$, $\mathcal{P_{\gamma \neq \gamma_{\rm i}}}$ is invariably zero indicating that all the fluctuations occur along the initial string. At finite $J$, fluctuations start to take place. As expected, these are significantly more visible in the unbroken case. We note that the string breaking is not significantly affected by the presence of a plaquette term as can be seen by the expectation value of the particle number $\langle \hat n \rangle$. In (c), we also show comparisons with the minimal model, which is strictly valid deep in the confined phase $g\rightarrow \infty$.
 }
 \label{fig:hex_z2}
\end{figure*}

We start by analyzing the fluctuations of unbroken strings in the confined phase and away from the first-order resonance $h=2m$. We consider a string that stretches along the boundary of the system, covering a path $\gamma_{\rm i}$ as shown in Fig.~\ref{fig:hex_z2}(c). The amount of oscillations to other string configurations of the same length can be measured by tracking both the fidelity $\mathcal{F}$ and the overlap $\sum_{\gamma}\mathcal{P}_{\gamma \neq \gamma{\rm i}}$ of the time evolved state with the other minimal strings. While drops in the fidelity can correspond to small matter fluctuations, which are significant at small $m$, a zero overlap with the other configurations shows that the string is immobile. In Fig.~\ref{fig:hex_z2}(a) we indeed observe that at $J=0$ there are no string oscillations over experimentally relevant timescales even as the electric field $g$ is decreased to small values. The explicit introduction of a small but finite plaquette term, on the other hand, introduces fluctuations over a timescale $t\approx1/J$. 

Similarly, at resonance, we observe that $\sum_{\gamma}\mathcal{P}_{\gamma \neq \gamma{\rm i}}$ is strictly zero at $J=0$. We note that while a finite plaquette term triggers string oscillations that coexist with the string broken states, this does not qualitatively affect the timescales of string-breaking and associated pair formation processes. 

\begin{figure*}
\centering
\includegraphics[width=0.8 \linewidth]{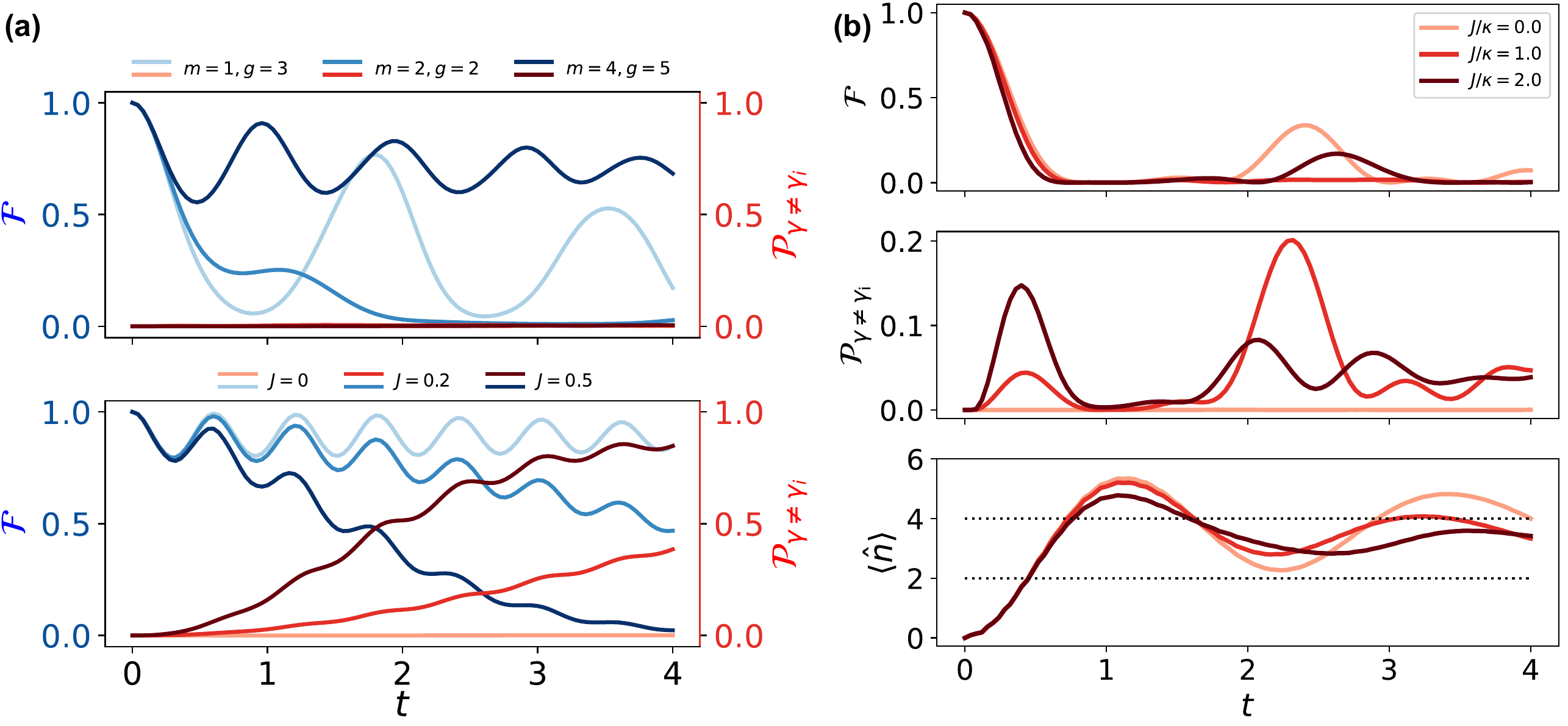}
\caption{Same as Fig.~\ref{fig:hex_z2}, but on $6 \times 4$ square lattice and taking as initial state an L-shaped string with $l_x$=4 and $l_y$=2.}
\label{fig:sq_z2_L}
\end{figure*}

\begin{figure*}
\centering
\includegraphics[width=0.8 \linewidth]{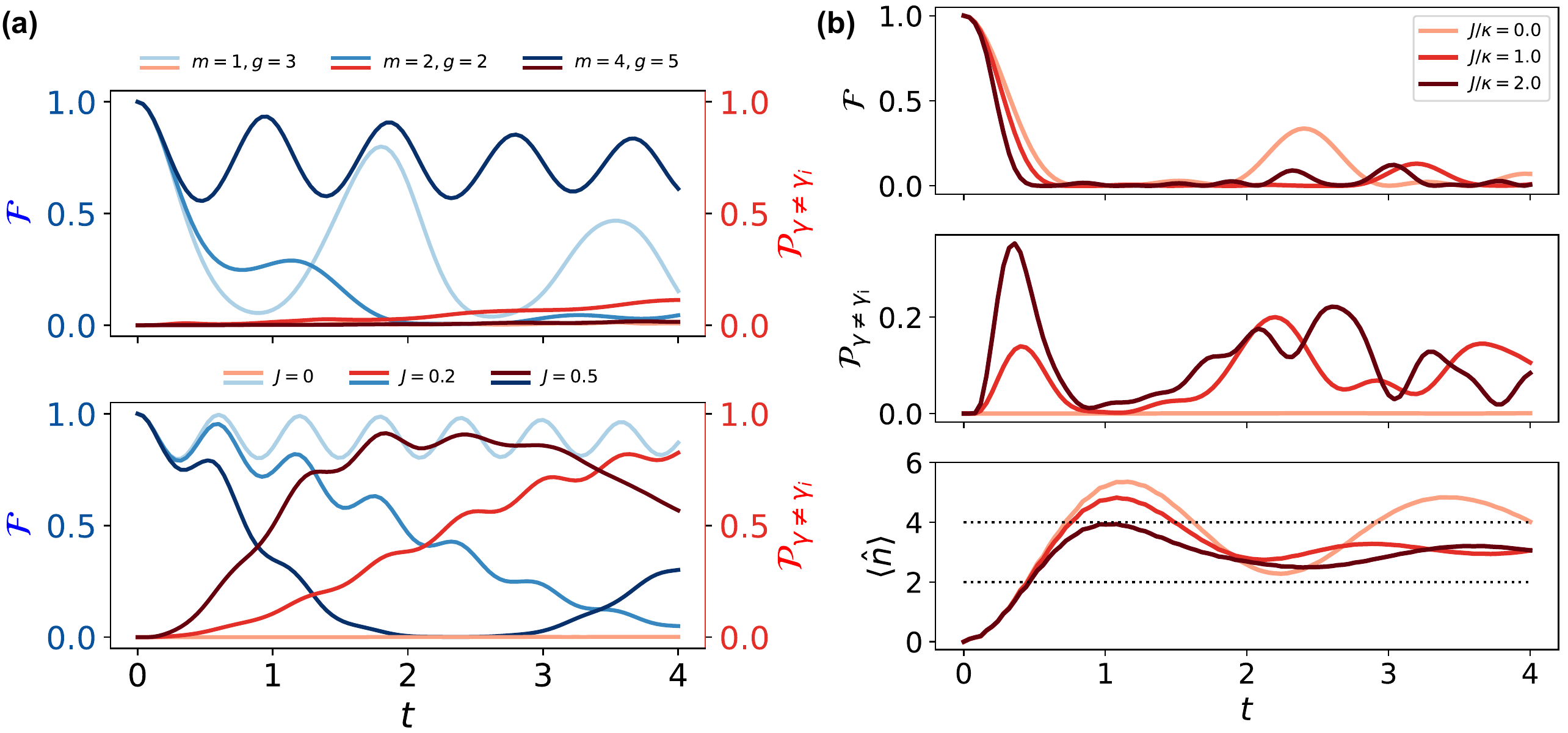}
\caption{Same as Fig.~\ref{fig:sq_z2_L}, but with a diagonal initial string.}
\label{fig:sq_z2_diag}
\end{figure*}

\subsection{Square lattice}
In Figs. \ref{fig:sq_z2_L} and \ref{fig:sq_z2_diag} we repeat the same computations of Fig.~\ref{fig:hex_z2} for a square lattice of size $6 \times 4$, starting with an L-shaped and staircase string respectively. While the results are largely the same and fully support the conclusions described above, we note the following interesting exception: in the upper panel of Fig.~\ref{fig:sq_z2_diag}, one can see that for the diagonal string, a very small $\mathcal{P}_{\gamma \neq \gamma_\text{i}}$ can be induced even without the plaquette term when the parameters are fine-tuned to the second order resonance $m=g$. 

Indeed, at this resonance length-two mesons can break the string along an L-shaped segment covering two adjacent sides of a plaquette, while a subsequent process can reconstruct the string by annihilating the newly created pair of $\Zt$ particles and flipping the electric field on the other two sides of the plaquette, thus completing the L-to-L move to the opposite orientation. While such processes are higher-order and energetically suppressed, they nonetheless allow local fluctuations in the string orientation, explaining our observation. Such fluctuations can occur whenever a string forms a corner, but the structure of the diagonal string, which is composed of a series of minimal L-shaped segments, strongly enhances it and makes it visible at relatively short timescales. We also note that, consistently, no such phenomenon occurs on the hexagonal lattice, since the same process would involve mesonic resonances of order three and is therefore heavily suppressed.

\section{Additional material on string dynamics for the $2+1$D Quantum link model on the square lattice}

\begin{figure*}[ht]
 \centering
 \includegraphics[width=12cm]{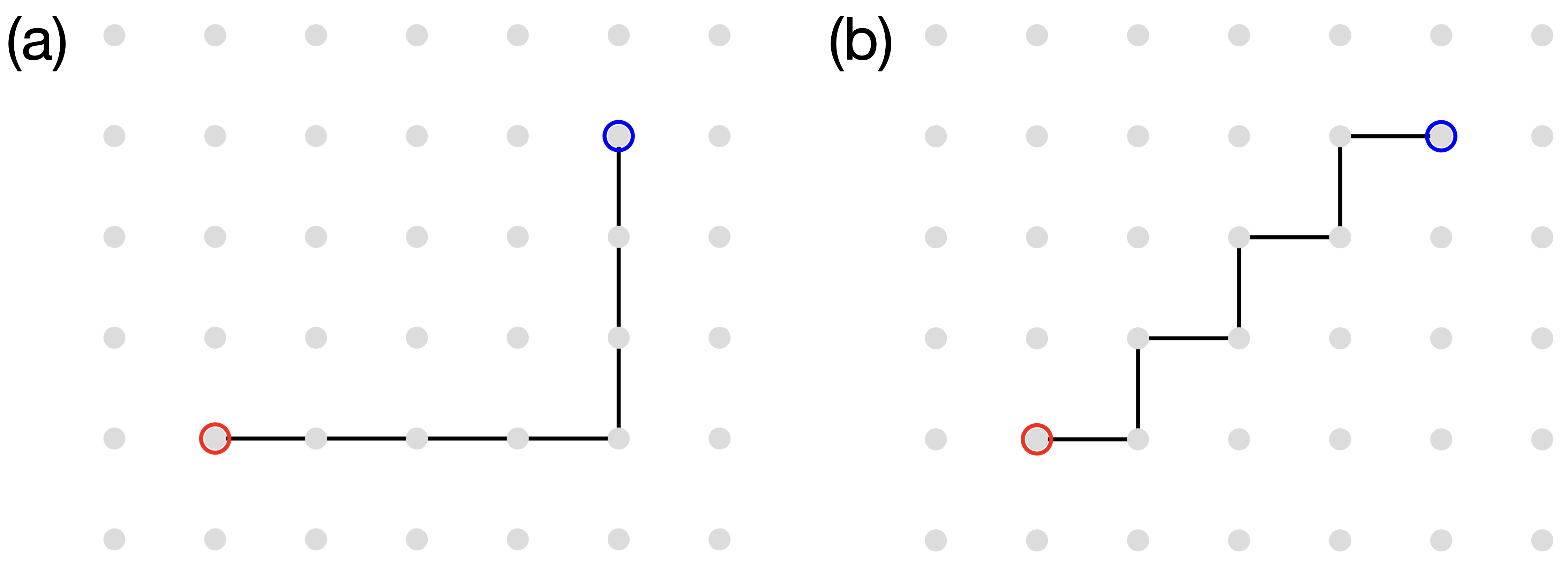}
 \caption{Depiction of the square lattice with string configurations connecting two static charges (blue and red circles). (a) L-shaped string studied in the main text. (b) Diagonal string considered here in the appendix.
 }
 \label{square_lat}
\end{figure*}

In this section, we present additional numerical details on the string dynamics considered for the QLM on the square lattice. We begin by presenting the lattice with the initial string configurations for the square lattice QLM studied in this work. While the main text focuses on the L-shaped string, here we also analyze the diagonal string, which is illustrated in Fig.~\ref{square_lat}.

\subsection{Effect of different initial string configurations on the string dynamics.}

\begin{figure*}[ht]

    \centering
    
    \includegraphics[width=0.49\textwidth]{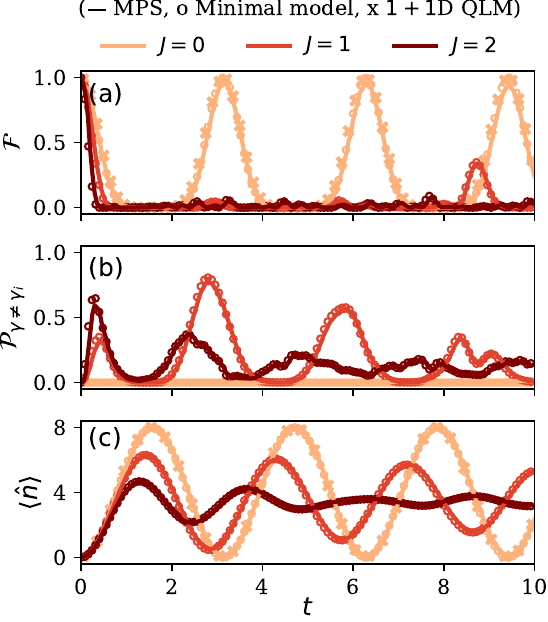}
    \includegraphics[width=0.49\linewidth]{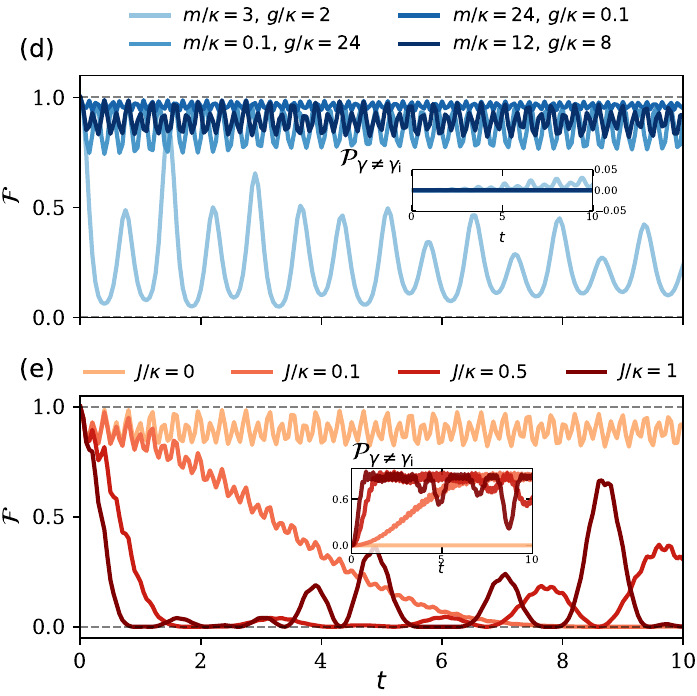}
      
    \caption{Dynamics of a diagonal string shown in Fig.~\ref{square_lat}. Left : Resonant string dynamics at $m/\kappa=12$, $g/\kappa=24$ and different values of $J/\kappa$, compared with the minimal model~$\hat{H}_\text{min}$. (a) The fidelity $\mathcal{F}$ with the initial string state. (b) The total overlap $\mathcal{P}_{\gamma\neq\gamma_{\text{i}}}$ with all minimal strings excluding the initial string. (c) The total matter occupation $\langle\hat{n}\rangle$ computed within the minimal patch containing the two static charges.  For $J/\kappa=0$, the $1+1$D QLM data is also shown for total matter occupation and the fidelity.  Right : The evolution of the initial-state fidelity following the time evolution of the diagonal string for different off-resonant parameters. The insets show the overlap  $\mathcal{P}_{\gamma\neq\gamma_{\text{i}}}$ between the time-evolved string and all other string configurations that have the same Manhattan distance as the diagonal string. (d) We fix $J/\kappa=0$ and choose different values for $m/\kappa$ and $g/\kappa$. (e) We fix $m/\kappa=12,\,g/\kappa=8$ and choose different values of $J/\kappa$.}\label{diagstr}

\end{figure*}

To investigate the effect of the initial string geometry on the resulting dynamics, we repeat the analysis of both resonant and off-resonant dynamics, as presented in the main text, but now starting from a diagonal initial string (see Fig.~\ref{square_lat}(b)). As shown in Fig.~\ref{diagstr}, we find no qualitative difference in the dynamics for the diagonal string compared to those observed for the L-shaped string. At resonance, the dynamics in the absence of plaquette terms ($J/\kappa = 0$) agree very well with both the minimal model and the $1+1$D QLM, as indicated by the fidelities $\mathcal{F}$, $\mathcal{P}_{\gamma\neq\gamma_{\text{i}}}$ (overlap between the time-evolved string and all other minimal string configurations) and matter occupation $\langle \hat{n} \rangle$.

We next numerically investigate the off-resonant dynamics to benchmark the essential role of plaquette terms in generating genuine $2+1$D string dynamics. As shown in Fig.~\ref{diagstr}, in the absence of plaquette interactions ($J/\kappa = 0$), the hopping process alone is insufficient to drive the system beyond its initial string configuration within the confined phase. This is evidenced by the sustained fidelity $\mathcal{F}$ and the vanishing overlap $\mathcal{P}_{\gamma \neq \gamma{\text{i}}}$ (see inset), indicating that the dynamics remain effectively $1+1$D. In contrast, when plaquette terms are included ($J/\kappa \neq 0$), we observe clear signatures of $2+1$D dynamics, characterized by a decaying fidelity and a nonzero $\mathcal{P}_{\gamma \neq \gamma{\text{i}}}$ (inset), reflecting nontrivial evolution away from the initial string. These results confirm that the emergence of $2+1$D string dynamics crucially depends on the presence of plaquette terms and is not sensitive to the specific choice of initial string geometry.

\subsection{Effect of small plaquette strength on the string dynamics.} 

\begin{figure*}[ht]
 \centering
 \includegraphics[width=8cm]{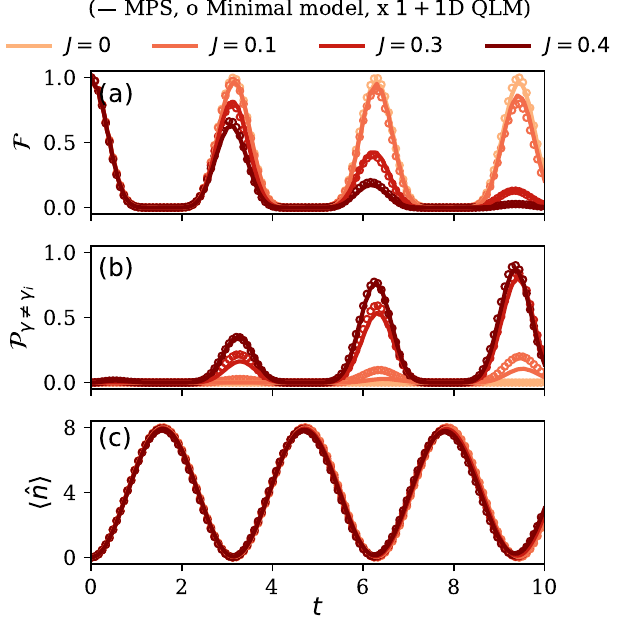}
 \caption{Resonant string dynamics for small values of $J/\kappa$ at $m/\kappa=12$, $g/\kappa=24$ starting from an L-shaped string computed using TDVP and the minimal model.  (a) The fidelity $\mathcal{F}$ with the initial string state. (b) The total overlap $\mathcal{P}_{\gamma \neq \gamma_{\text{i}}}$ with all minimal strings excluding the initial string. (c) The total matter occupation $\langle\hat{n}\rangle$ computed within the minimal patch containing the two static charges.}
 \label{smallJ}
\end{figure*}

It is important to understand the time scales associated with the onset of $2+1$D string dynamics induced by small plaquette terms. Intuitively, we expect such dynamics to emerge on a time scale $t \propto 1/J$, where $J$ is the strength of the plaquette terms. To test this intuition, we study string dynamics at resonance for small values of $J/\kappa$, as shown in Fig.~\ref{smallJ}. While the total matter occupation $\langle \hat{n} \rangle$ within the minimal patch shows revivals for all plaquette strengths considered, the fidelity and $\mathcal{P}_{\gamma \neq \gamma_{\text{i}}}$ (overlap with all minimal strings excluding the initial string) exhibit noticeable deviations. These deviations are consistent with our expectation that the emergence of true $2+1$D string dynamics occurs on a time scale that scales inversely with the plaquette strength.

\section{Quantum link model on the hexagonal geometry}
\begin{figure*}[ht]
 \centering
 \includegraphics[width=11cm]{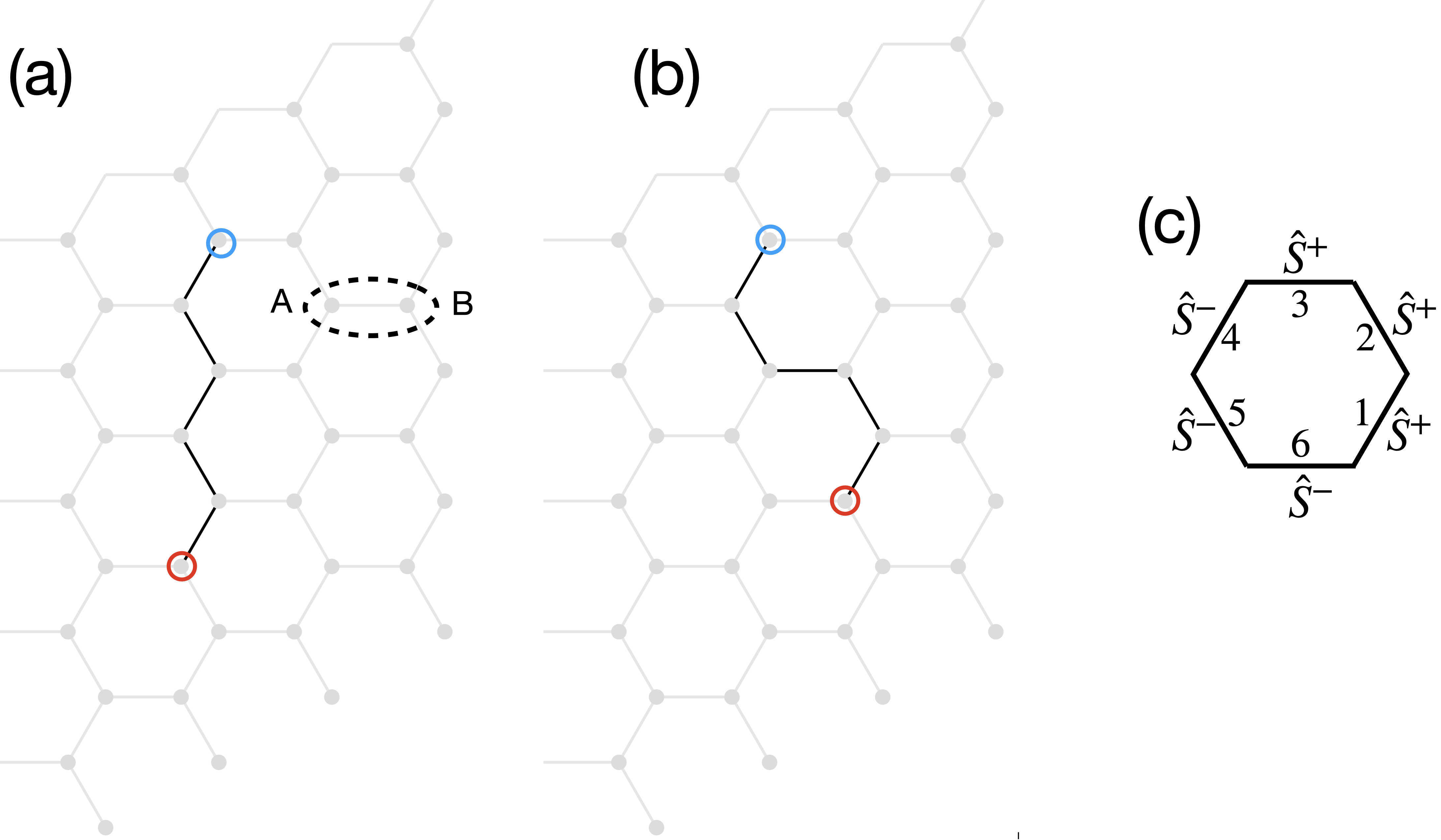}
 \caption{(a) Hexagonal lattice set up showing a minimal string of length $L=5$ between two static charges as in the \texttt{QuEra} \cite{gonzalezcuadra2024observationstringbreaking2} éxperiment. We use this an initial state to study string dynamics here in the appendix. (b) A S-shaped string state of the same minimal distance ($L=5$) but with two flippable plaquettes. We used this as an initial state to study string dynamics in Fig.\ref{reson_hex} in the main text. We consider this geometry on a cylinder with periodicity along the $y$-direction. (c) Schematic representation of the six-body plaquette interaction term.
 }\label{hex_qlm}
\end{figure*}

We investigate a quantum link model implemented on a hexagonal lattice, as realized in recent Rydberg atom experiments by \texttt{QuEra} \cite{gonzalezcuadra2024observationstringbreaking2}. The effective Hamiltonian realizing the $U(1)$ lattice gauge theory within the Rydberg atom array framework is given by

\begin{equation}
\begin{aligned}
    H_{\mathrm{eff}} = &\ -\kappa \sum_{\langle x, y \rangle} \left( \hat{\phi}_x^\dagger \hat{S}_{\langle x,y\rangle}^{+} \hat{\phi}_y + \mathrm{H.c.} \right) 
    + m \sum_x (-1)^{s_x} \hat{\phi}_x^\dagger \hat{\phi}_x \\
    & + g \sum_{\langle x, y \rangle} \hat{S}_{\langle x, y \rangle}^z - J \sum_{\Box}(\hat{S}^+_1\hat{S}^+_2\hat{S}^+_3\hat{S}^-_4\hat{S}^-_5\hat{S}^-_6 +\mathrm{H.c.}).
    \end{aligned}
\label{eq:hamiltonian-hex}
\end{equation}
The first term represents the minimal coupling between matter and gauge fields. The second term, which is staggered, sets the mass of the matter fields. The third term introduces a linear electric field, effectively generating string tension in the model (although the \texttt{QuEra} experiment \cite{gonzalezcuadra2024observationstringbreaking2} includes a long-range interaction, it effectively reduces to this linear form). The final term corresponds to the six-body plaquette interaction, as described in  Fig \ref{hex_qlm}(c). The Gauss's law operator at site $x$ in this geometry reads as,

\begin{equation}
    \hat{G}_x = \nabla \cdot \hat{\mathbf{S}}_x^z - \hat{Q}_x = 0,
    \label{eq:gauss}
\end{equation}
where $\nabla \cdot \hat{\mathbf{S}}_x^z$ represents the divergence of the electric field at site \( x \), constructed from spin-$1/2$ operators on the adjacent links, and \( \hat{Q}_x \) denotes the local dynamical charge given by $\hat{Q}_x=\hat{\phi}^{\dagger}_x\hat{\phi}_x-[1-(-1)^{s_x}]/2$. The allowed physical states \( |\Psi\rangle \) satisfy \( \hat{G}_x |\Psi\rangle = q_x |\Psi\rangle \), with a staggered background charge \( q_x = (-1)^{s_x}/2 \), where \( s_x \) is a site-dependent sign factor which is $+1$ for sites belonging to site A in the unit cell and $-1$ for sites belonging to site B.

\begin{figure*}[ht]
 \centering
 \includegraphics[width=9cm]{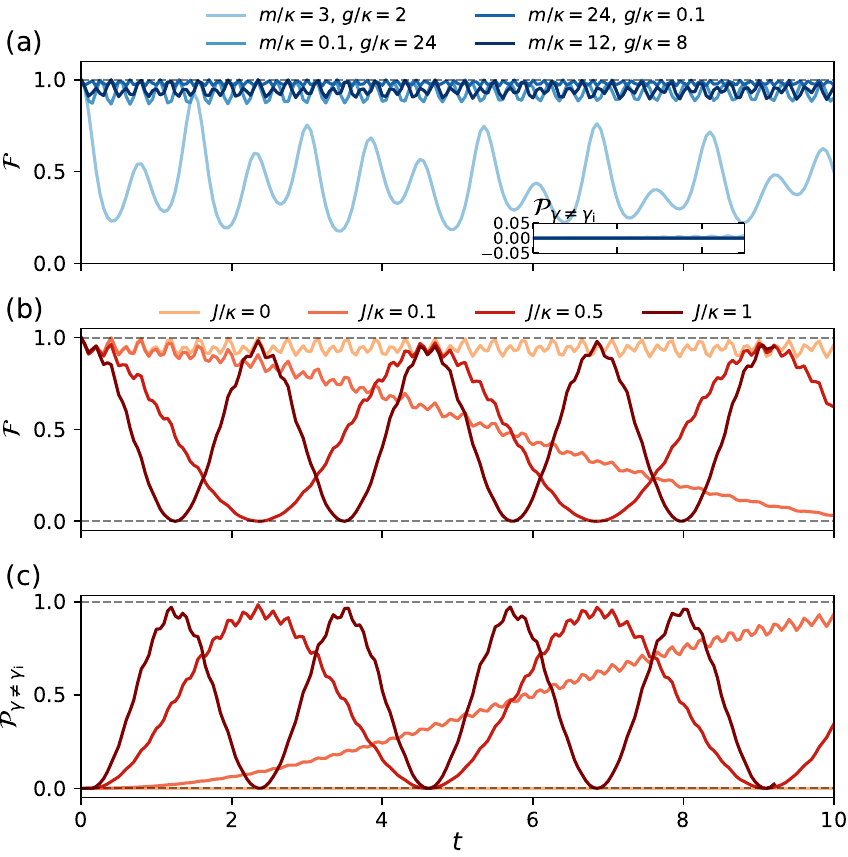}
 \caption{The fidelities computed for the initial string shown in Fig~\ref{hex_qlm} (b) for different parameters for the QLM on the hexagonal lattice. The inset shows the overlap  $\mathcal{P}_{\gamma\neq\gamma_{\text{i}}}$ between the time-evolved string and other string configurations that can be deformed from the original string by applying the plaquette term. (a) We fix $J/\kappa=0$ and choose different values for $m/\kappa$ and $g/\kappa$. (b) We fix $m/\kappa=12,\,g/\kappa=8$ and choose different values of $J/\kappa$. (c) $\mathcal{P}_{\gamma\neq\gamma_{\text{i}}}$ sfor exactly the same parameters as in (b).}
 \label{hex_off_resonance}
\end{figure*}

\subsection{Off-resonant dynamics}

In this section, we examine the off-resonant dynamics of the string configuration shown in Fig.~\ref{hex_qlm}(b). Following the approach used for the square lattice in the main text, we set the plaquette term to zero ($J/\kappa = 0$) and choose parameters away from resonance to suppress unwanted string breaking. Our goal is to determine whether the string evolves into a configuration distinct from its initial state. As shown in Fig.~\ref{hex_off_resonance}(a), in the absence of plaquette terms, hopping alone is \emph{insufficient} to drive string dynamics beyond the original configuration in the confined phase. This is reflected by the sustained fidelity $\mathcal{F}$ and the vanishing probability $\mathcal{P}_{\gamma \neq \gamma_{\text{i}}}$ (shown in the inset). When the plaquette terms are turned on ($J/\kappa \neq 0$), Fig.~\ref{hex_off_resonance}(b,c) shows that the dynamics become genuinely $2+1$D. This is evidenced by a decaying fidelity for $J/\kappa = 0.1$ and a corresponding nonzero value of $\mathcal{P}_{\gamma \neq \gamma_{\text{i}}}$. For larger values of $J/\kappa$, we observe oscillations in both the fidelity $\mathcal{F}$ and $\mathcal{P}_{\gamma \neq \gamma_{\text{i}}}$. These oscillations arise from plaquette-only dynamics, which induce coherent revivals among the three possible minimal string configurations, resembling the behavior of a three-level quantum system.

\subsection{Resonant dynamics of the $1$d initial string}

\begin{figure*}[ht]
 \centering
 \includegraphics[width=8cm]{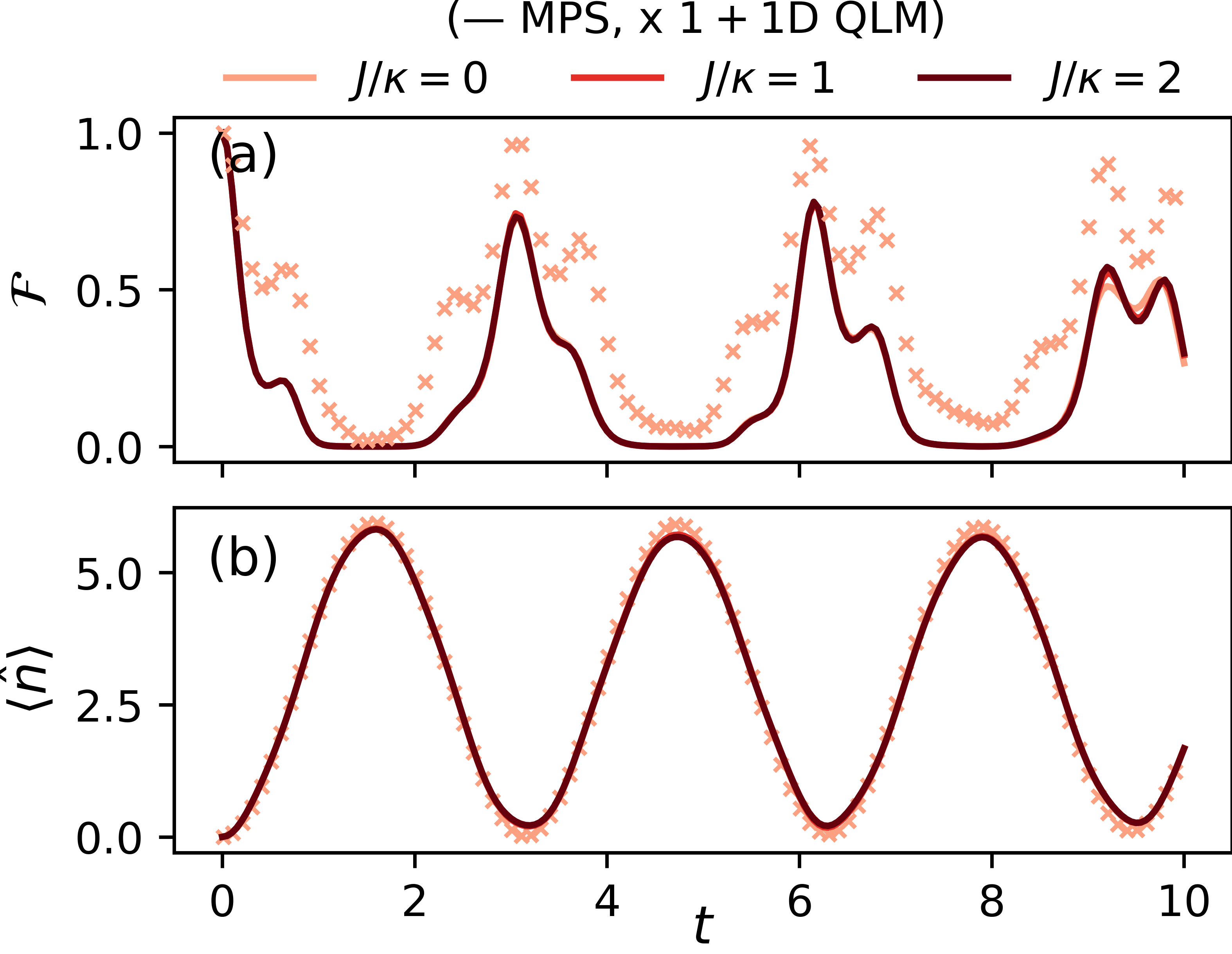}
 \caption{Resonant string dynamics for a simple $1$d string shown in Fig.~\ref{hex_qlm}(b) for $m/\kappa=2,g/\kappa=4$ and for different values of $J/\kappa$. (a) Fidility $\mathcal F$ with the initial state. (b) The total matter occupation $\langle \hat{n}\rangle$ on the string. For $J/\kappa=0$, the $1+1$D QLM data is also shown for the total matter occupation and the fidelity. }
 \label{1dstr}
 \end{figure*}

 Here, we study the resonant dynamics of a $1$d string configuration, also considered in the experiments by \texttt{QuEra}~\cite{gonzalezcuadra2024observationstringbreaking2}, and shown in Fig.~\ref{hex_qlm}(b). We choose the same resonance condition as in the experiment, setting $2m = g$ with $m/\kappa = 2$ and $g/\kappa = 4$. As illustrated in Fig.~\ref{1dstr}, we observe near-perfect revivals in both the matter occupation and the fidelities, regardless of whether plaquette terms are included. This behavior is due to the absence of any flippable plaquettes in this simple $1$d string configuration. These results support the central message of our work that hopping terms alone are insufficient to generate genuine $2+1$D string dynamics.

\begin{figure*}[ht]
 \centering
 \includegraphics[width=0.9\textwidth]{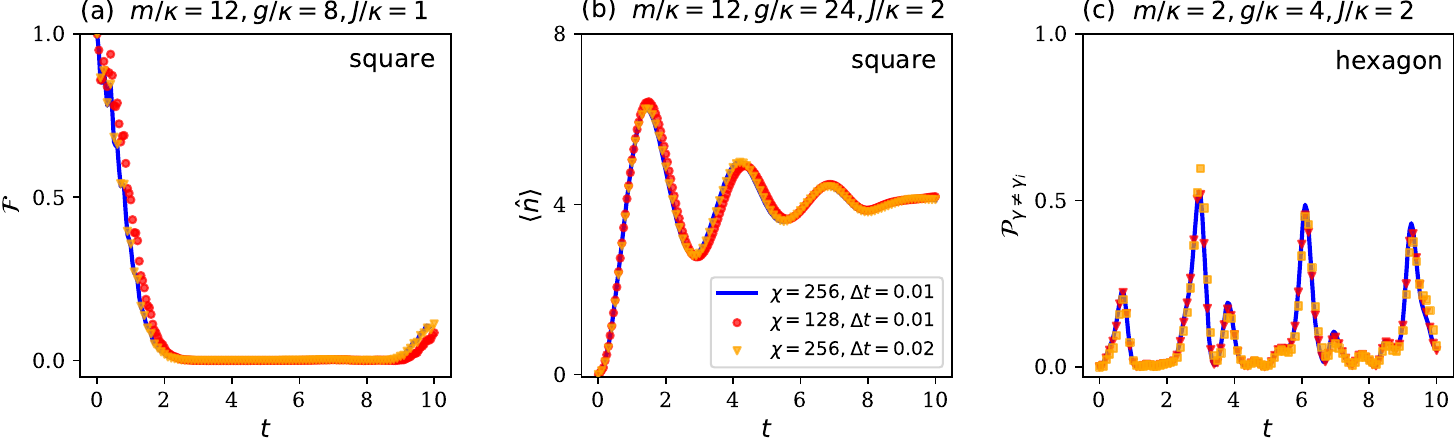}
 \caption{Convergence analysis of  observables for different bond dimensions $\chi$ and TDVP time steps $\Delta t$, evaluated across various parameter regimes for string dynamics in the QLM. The study is performed on the square lattice with an L-shaped initial string and on the hexagonal lattice with the S-shaped initial string, both considered in the main text. (a) Fidelity of the initial L-shaped string on the square lattice. (b) Total matter occupation within the patch for the square lattice. (c) Overlap $\mathcal{P}_{\gamma \neq \gamma_{\text{i}}}$ between the time-evolved state and all string configurations excluding the initial minimal string on the hexagonal lattice.}
 \label{convergence}
 \end{figure*}

\section{Convergence tests}

We numerically study the real-time dynamics of string configurations using the time-dependent variational principle (TDVP) algorithm~\cite{Haegeman2011,Haegeman2013,Haegeman2016}, implemented in the \texttt{Matrix Product Toolkit}~\cite{mptoolkit}. To ensure the accuracy of our simulations, we perform a detailed convergence analysis by varying both the bond dimension $\chi$ and the TDVP time step $\Delta t$ as shown in Fig.~\ref{convergence}. In particular, we test convergence across different parameter regimes of the QLM, with a focus on regions where entanglement growth is expected to be significant and where higher bond dimensions are required to faithfully capture the dynamics. Our analysis includes simulations on both the square lattice, initialized with an L-shaped string, and the hexagonal lattice, initialized with the S-shaped string geometries studied in the main text. These benchmarks validate that the numerical results presented in this work are converged in the simulation parameters used to a good accuracy.

\section{Bottlenecks in defining a string in the spin-$1/2$ QLM} \label{qlmstrings}

Here, we discuss possible string configurations between two static charges in the spin-$\frac{1}{2}$ QLM. One possible vacuum configuration in this model in the physical sector of Gauss's law $\hat{G}_{\mathbf{j}}|\Psi\rangle = 0$ corresponds to a state where all spins point either to the left or downward (i.e., with magnitude $-\frac{1}{2}$), as illustrated in Fig.~\ref{qlmstrings}(b). A bottleneck arises when defining string states between two static charges. Specifically, only strings of \textit{minimal length} are allowed in this model (defining length using the Manhattan distance). Any non-minimal string would necessarily bend around corners, which leads to violations of Gauss's law. For example, in the simple case depicted in Fig.~\ref{qlmstrings}(d), the \textit{snake} string violates Gauss’s law at the corners marked 1, 2, and 3, as $\hat{G}_{\mathbf{j}}|\Psi\rangle \neq 0$ on those sites. The deeper reason for this restriction lies in the nature of the allowed vacuum configurations (see Fig.~\ref{qlmstrings}(a)), which are constrained by the half-integer representation of the gauge fields. This representation restricts the types of electric fluxes and hence the permissible string configurations. We also note that this restriction also applies to the QLM on the hexagonal geometry considered in this work.

\begin{figure*}[ht]
 \centering
 \includegraphics[width=10cm]{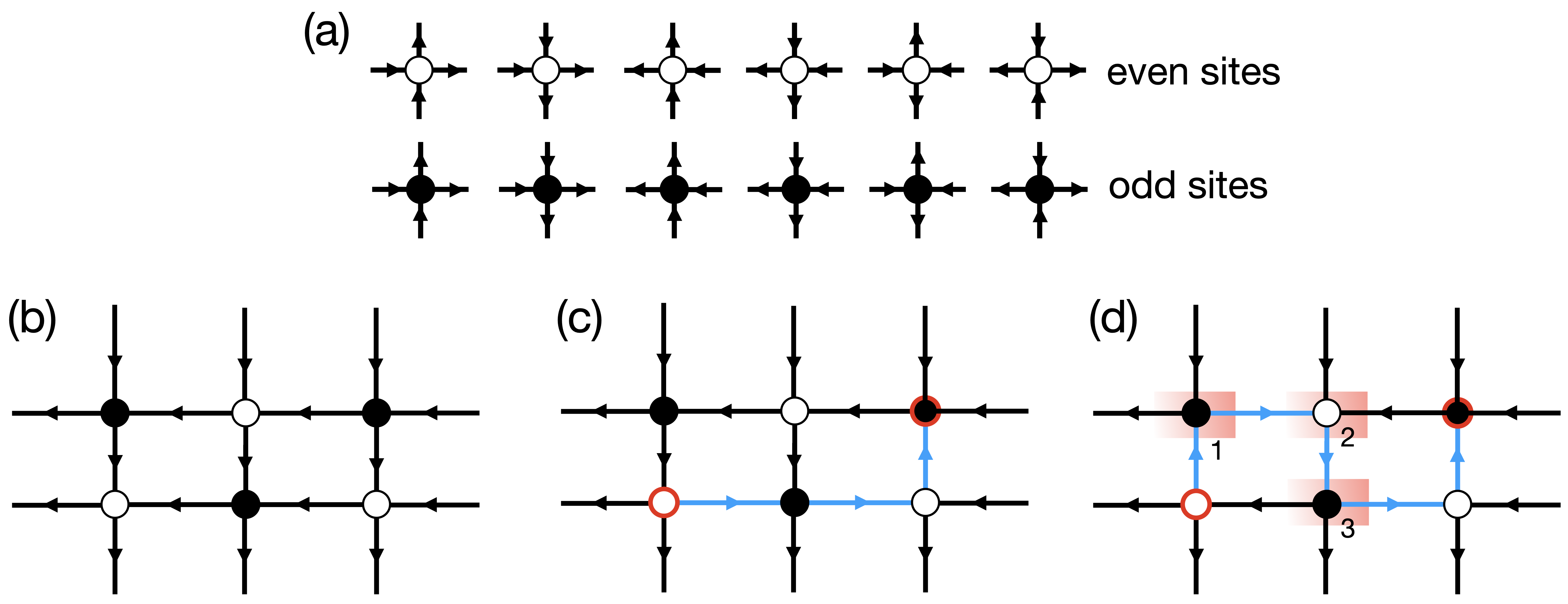}
 \caption{(a) The charge-neutral vacuum gauge-site configurations allowed by Gauss's law on odd and even sites. (b) A gauge-site configuration for the vacuum product state. (c) A minimal string (blue) between two static charges (circled red) (c) A non-minimal snake shaped string (blue) that violates Gauss's law at the corners (1,2,3) where the string bends.}
 \label{qlmstrings}
 \end{figure*}

\section{Elaboration on the minimal model construction}

\begin{figure*}[ht]
 \centering
 \includegraphics[width=10cm]{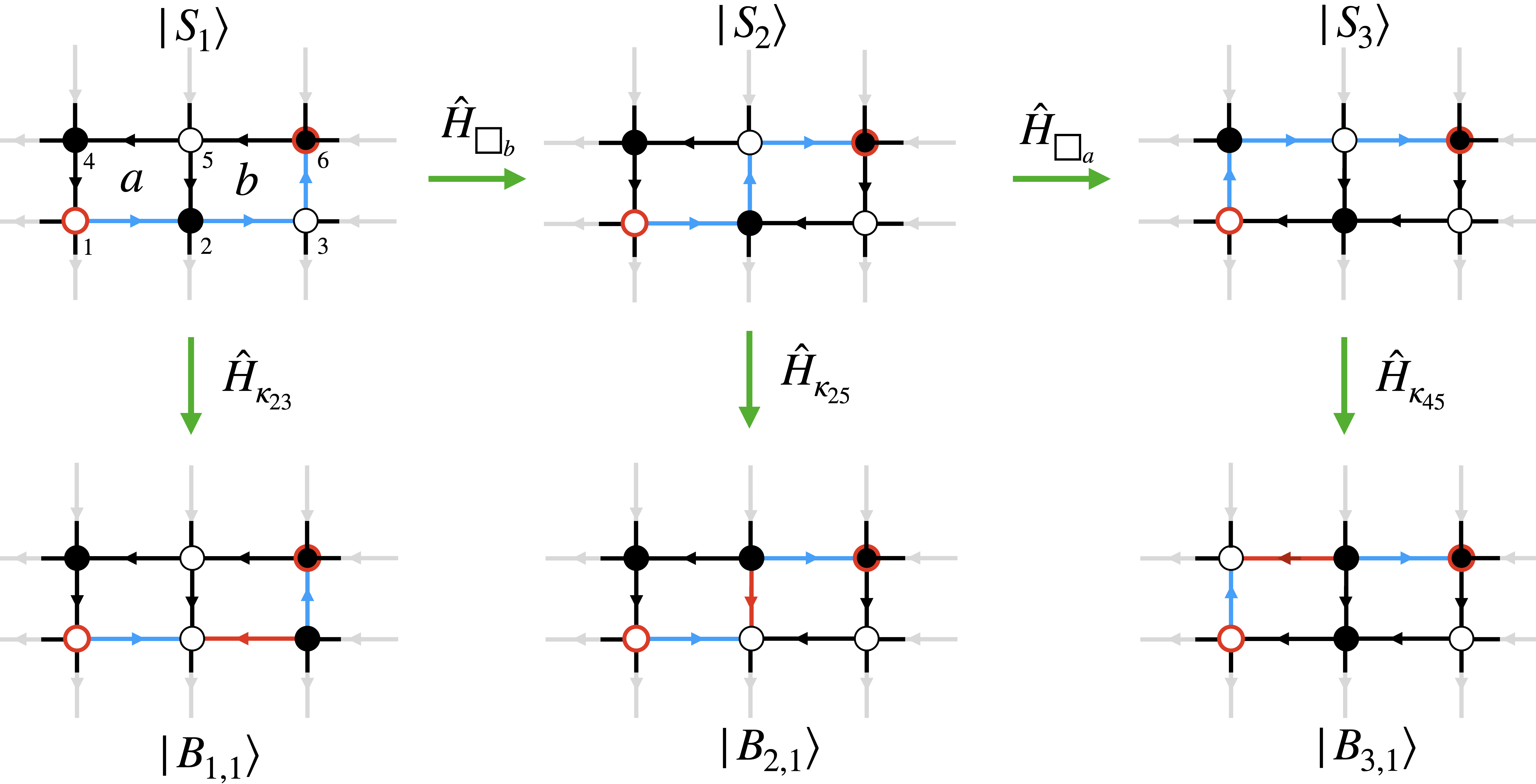}
 \caption{An illustration showing how the fixed-energy manifold is built from configurations with broken and unbroken strings using plaquette and hopping operations.}
 \label{minmod}
\end{figure*}

The idea behind the minimal model is to capture string dynamics at first-order resonance, where the condition $2m = g$ holds, and in the deeply confinemed regime characterized by $2m + g \gg \kappa$. In this regime, the Hilbert space simplifies significantly, consisting primarily of \textit{intact string} and \textit{broken string} states. The intact string states denoted as $|S_n\rangle$ can be generated by repeated action of the plaquette operator on an initial state, such as $|S_1\rangle$, wherever applicable (a plaquette term can only act on a flippable plaquette available within the minimal patch $\mathcal{S}$). That is,
\begin{equation}
|S_n\rangle = \left( \sum_{\Box\in \mathcal{S}}\hat{H}_{\Box} \right)^n |S_1\rangle,
\end{equation}
which generates a closed set of minimal string states $|S_n\rangle$, all with equal energy. Similarly, starting from an intact string state $|S_i\rangle$, one can construct the set of all corresponding broken string states $|B_{i,l}\rangle$ by repeatedly applying the hopping term, acting only on the lattice sites $l$ that belong to the string $S_i$. Importantly, only configurations that conserve energy needs to kept. That is,
\begin{equation}
|B_{i,l}\rangle = \left( \sum_{l \in S_i} \hat{H}_{\kappa, (l, l+1)} \right)^n_{E=\textrm{fixed}} |S_i\rangle.
\end{equation}
We then project the full Hamiltonian onto this equal-energy manifold using the projection operator $\hat{P}$, resulting in an effective Hamiltonian of the form,
\begin{equation}
\hat{H}_{\text{min}} = \hat{P} \hat{H} \hat{P}.
\end{equation}
A schematic example illustrating the construction of this projected subspace for a string consisting of three links is shown in Fig.~\ref{minmod}.

For the case analyzed in the main text, involving a string of length $L=7$ on a $5\times 4$ region of the square lattice, the dimension of the projected subspace is $\mathcal{D}=560$. To study the string dynamics, we exponentiate the resulting matrix numerically.

\end{document}